\documentclass{mem}
\usepackage{natbib}\usepackage{txfonts}\usepackage{balance}
\usepackage{graphicx}
\usepackage[a4paper,breaklinks,dvipdfm]{hyperref}
\idline{83}{282}

\newcommand{\COBOLD}{{\tt CO$^5$BOLD}}
\newcommand{\Stagger}{{\tt Stagger}}
\newcommand{\MARCS}{{\tt MARCS}}

\newcommand{\moh}{\ensuremath{[\mathrm{M/H}]}}
\newcommand{\Teff}{\ensuremath{T_{\mathrm{eff}}}}
\newcommand{\tauross}{\ensuremath{\tau_{\mathrm{Ross}}}}
\newcommand{\xtmean}[1]{\ensuremath{\left\langle #1\right\rangle}}
\newcommand{\mlp}{\ensuremath{\alpha_{\mathrm{MLT}}}}

\begin{document}
\def\teff{$T\rm_{eff }$}
\def\kms{$\mathrm {km s}^{-1}$}

\title{
The influence of convection on the atmospheric structures and observable properties of red giant stars}

   \subtitle{}

\author{
        A. \,Ku\v{c}inskas\inst{1,2}
        \and
        H.-G. \,Ludwig\inst{3}
        \and
        M. \,Steffen\inst{4}
        \and
        V. \,Dobrovolskas\inst{2,1}
        \and
        J. \,Klevas\inst{1}
        \and\\
        D. \,Prakapavi\v{c}ius\inst{1}
        \and
        E. \,Caffau\inst{3}
        \and
        P. \,Bonifacio\inst{5}
       }

  \offprints{A. \,Ku\v{c}inskas}

\institute{
        Vilnius University Institute of Theoretical Physics and Astronomy, A. Go\v {s}tauto 12, Vilnius LT-01108, Lithuania;
        \email{arunas.kucinskas@tfai.vu.lt}
        \and
        Vilnius University Astronomical Observatory, M. K. \v{C}iurlionio 29, Vilnius LT-03100, Lithuania
        \and
        ZAH Landessternwarte K\"{o}nigstuhl, D-69117 Heidelberg, Germany
        \and
        Leibniz-Institut f\"ur Astrophysik Potsdam, An der Sternwarte 16, D-14482 Potsdam, Germany
        \and
        GEPI, Observatoire de Paris, CNRS, Universit\'{e} Paris Diderot, Place Jules Janssen, 92190 Meudon, France
}

\authorrunning{Ku\v{c}inskas et al.}

\titlerunning{Convection and observable properties of red giant stars}

\abstract{
During the recent years significant progress has been made in the modeling of red giant atmospheres with the aid of 3D hydrodynamical model atmosphere codes. In this contribution we provide an overview of selected results obtained in this context by utilizing 3D hydrodynamical \COBOLD\ stellar model atmospheres. Hydrodynamical simulations show that convective motions lead to significant differences in the atmospheric structures of red giants with respect to those predicted by the classical 1D model atmospheres. Results of these simulations also show that in certain cases 1D models fail to reproduce even the average properties of the 3D hydrodynamical models, such as $P-T$ profiles. Large horizontal temperature fluctuations in the 3D model atmospheres, as well as differences between the temperature profiles of the average $\xtmean{\mbox{3D}}$ and 1D models, lead to large discrepancies in the strengths of spectral lines predicted by the 3D and 1D model atmospheres. This is especially important in models at lowest metallicities ($\moh<-2.0$) where the ${\rm 3D-1D}$ abundance differences may reach (or even exceed) $-0.6$\,dex for lines of neutral atoms and molecules. We also discuss several simplifications and numerical aspects involved in the present 3D hydrodynamical modeling of red giant atmospheres, and briefly address several issues where urgent progress may be needed.
}
\maketitle{}

\section{Introduction}

Red giant stars are commonly present in all intermediate age and old stellar populations. Because of their high intrinsic luminosity, they are amongst the few classes of objects accessible for study in remote stellar populations, or populations that are heavily obscured by interstellar extinction. This makes them attractive and useful tracers of stellar populations in the Galaxy and beyond.

The observable properties of red giant stars, however, are still relatively poorly understood. In part, this is because until now their atmospheres have been routinely studied with the aid of one-dimensional (1D) stationary model atmospheres, which have to rely on a number of simplifying assumptions and free parameters. Shortcomings of the stationary 1D models are especially evident in the context of modeling atmospheric convection, which is inherently a three-dimensional (3D) time-dependent phenomenon and therefore the usefulness of classical 1D model atmospheres in this context is limited.

To partly fill in this gap, we have recently started a project to study the influence of convection on the atmospheric structures and observable properties of red giant stars, by focusing on the dynamical properties of their atmospheres, spectral line formation, emergent spectral energy distributions, and photometric colors \citep[for first results see, e.g.,][]{KLC09,DKL10,IKL10,LK12,KSL13,DKS13,KKL13,PSK13}. This work is carried out using state-of-the-art 3D hydrodynamical \COBOLD\ model atmosphere package \citep[][]{FSL12}. In this contribution we briefly summarize some of the early results obtained in the course of this project, and discuss selected problems that need to be solved in order to make further progress in the modeling of red giant atmospheres and understanding their observable properties with the aid of 3D hydrodynamical model atmospheres.

\begin{figure*}[t!]
\resizebox{\hsize}{!}{\includegraphics[width=6cm]{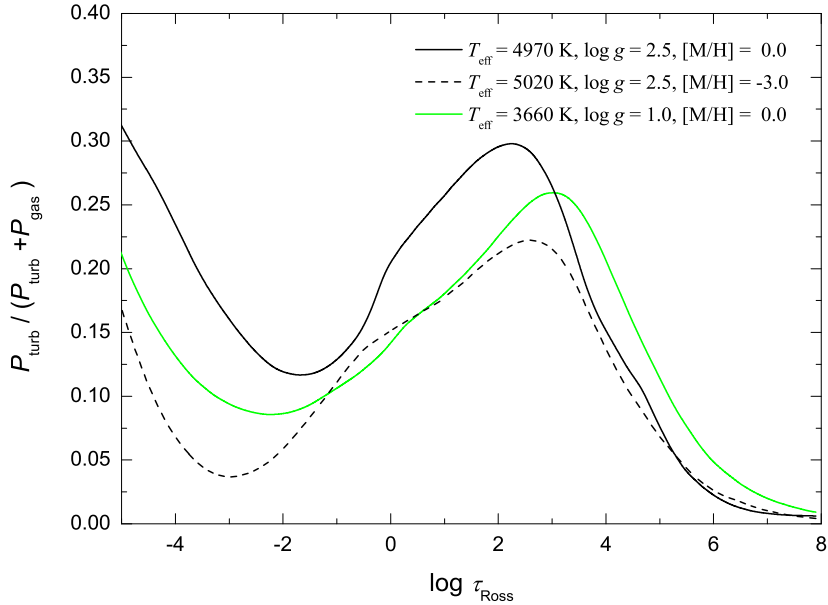}}
\caption{\footnotesize
Average ratio of the turbulent pressure, $P_{\rm turb}$, to the total pressure, $P_{\rm turb}+P_{\rm gas}$, as a function of Rosseland optical depth in the 3D hydrodynamical models of red giants located close to the tip ($\Teff\approx3660$\,K, $\log g=1.0$, $\moh=0.0$) and bottom ($\Teff\approx5000$\,K, $\log g=2.5$, $\moh=0.0$ and $\moh=-3.0$) of the RGB. In all cases, pressure ratios were calculated on surfaces of equal Rosseland optical depth using one 3D model snapshot (i.e., 3D model structure obtained at one particular instant in time).
}
\label{fig:pturb}
\end{figure*}

\section{Convection and observable properties of red giant stars}

\subsection{Convective properties of red giant atmospheres\label{sect:struct}}

Convective motions in the atmospheres of red giant stars lead to the formation of surface granulation, akin to the one observed on the surface of the Sun. Qualitatively, the properties of granulation pattern predicted by the 3D hydrodynamical models of the Sun and red giants are very similar: flows of uprising hot matter are surrounded by narrower intergranular lanes that are formed by flows of cooler matter directed downwards into the stellar interiors. The detailed structure of the convective pattern is, however, quite different.

One notable difference is that granules have significantly larger relative size in red giants than they do in the Sun. With a typical granule size of $\sim1$\,Mm in the Sun, there are $\sim10^6$ granules observable at any time on the surface of the Sun. The corresponding number is significantly lower in giants: the size of granules may range from $\sim250$\,Mm at $\Teff\approx5050$\,K, $\log g=2.5$, and $\moh=0.0$, to $\sim5$\,Gm at $\Teff\approx3660$\,K, $\log g=1.0$, and $\moh=0.0$, or, correspondingly, to $\sim1000$ and $\sim400$ granules observable on the stellar surface \citep[][]{KSL13,DKS13}. Both vertical and horizontal velocities of the convective flows seem to be higher in the giants as well: they may reach up to $\sim2.5$ and $\sim6$\,Mach in the atmosphere of a giant, respectively ($\Teff\approx3660$\,K, $\log g=1.0$, $\moh=0.0$), whereas they typically do not exceed 1.5 and 1.8\,Mach in the Sun \citep[][]{LK12}.

Because of the strong convective flows in red giant atmospheres, often even the average properties of the 3D hydrodynamical models can not be satisfactorily reproduced with the classical 1D model atmospheres. For example, analysis of the 3D hydrodynamical \COBOLD\ model of a red giant with $\Teff\approx3660$\,K, $\log g=1.0$ and $\moh=0.0$ shows that turbulent pressure may play significantly more important role in giants than it does in the Sun. Turbulent pressure alters the $P-T$ relation in the average $\xtmean{\mbox{3D}}$ hydrodynamical model to such an extent that it can not be reproduced with the 1D model atmospheres, despite any chosen value of the turbulent pressure factor and/or the mixing-length parameter \citep[see][for details]{LK12}.

The contribution of the turbulent pressure to the total pressure is significant in giants with higher effective temperatures and gravities too. This is clearly seen in Fig.~\ref{fig:pturb} where we plot $P_{\rm turb}/(P_{\rm turb}+P_{\rm gas})$ versus optical depth in the atmospheres of giants located near the tip and bottom of the red giant branch (RGB), at two metallicities ($\moh=0.0$ and $\moh=-3.0$) in the latter case. One may notice that $P_{\rm turb}/(P_{\rm turb}+P_{\rm gas})$ is double-peaked, increasing both towards the deeper atmospheric layers below the optical surface ($\log \tauross>0$) and the outer atmosphere. The turbulent pressure is somewhat lower at the optical depths where the majority of spectral lines form (e.g., $\log\tauross\sim-3.0\dots0.0$) but even in these regions it contributes $\sim5-15$\% to the total pressure. While the importance of turbulent pressure decreases towards lower metallicity (especially in the outer atmosphere), it remains non-negligible even in the lowest metallicity giant at $\moh=-3.0$ located near the bottom of the RGB (Fig.~\ref{fig:pturb}).

\begin{figure*}[t!]
\resizebox{\hsize}{!}{\includegraphics[clip=true]{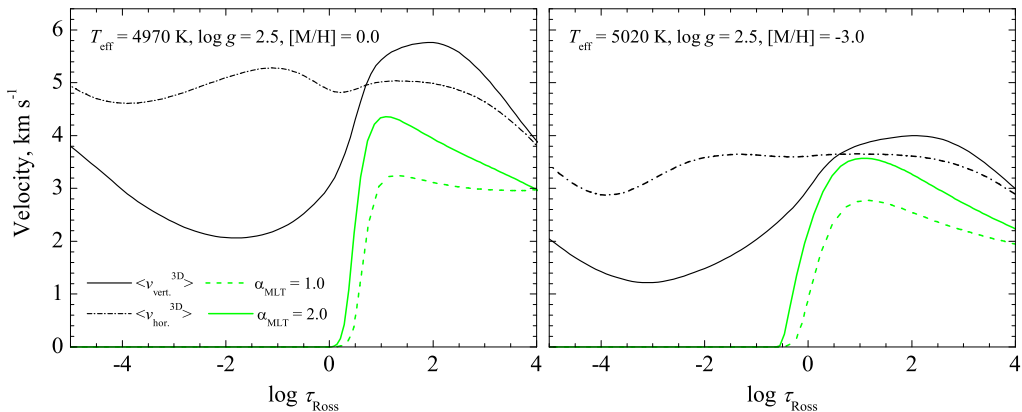}}
\caption{\footnotesize
Velocity profiles in the 3D hydrodynamical and classical 1D model atmospheres of a red giant ($\Teff\approx5000$\,K, $\log g=2.5$), at $\moh=0.0$ (left) and $\moh=-3.0$ (right). Black solid and dot-dashed lines are correspondingly the average vertical and horizontal RMS velocities in the 3D hydrodynamical model calculated using twenty 3D model snapshots averaged on surfaces of equal Rosseland optical depths. Grey (green) solid and dashed lines are convective velocities in the 1D models calculated according to the mixing-length theory in the formulation of Mihalas \citep[see][]{LFS99} with two mixing length parameters, $\mlp=1.0$ and $\mlp=2.0$.
}
\label{fig:vel-prof}
\end{figure*}

In the giant model studied by \citet[][]{LK12}, ${\rm 3D-1D}$ changes in the $P-T$ profile occur significantly below the optical surface an thus have little direct influence on the observable properties of the star. One should note, however, that convective motions tend to reach to lower optical depths at lower metallicities than they do at $\moh=0$, thus turbulent pressure may force changes in the observable properties of giants at $\moh<0$. This is illustrated in Fig.~\ref{fig:vel-prof} where we show convective velocity profiles in the 3D hydrodynamical and 1D classical model atmospheres at $\moh=0.0$ and $\moh=-3.0$. Clearly, both 3D and 1D models predict that convective motions should reach farther into the atmosphere at low metallicity and may thus directly alter the $P-T$ profiles in the regions where spectral lines form. Even more importantly, 3D hydrodynamical models predict significant overshoot of matter into the regions that should be convectively stable according to the classical Schwarzschild criterion. This results in the vertical velocity profiles that are very different from those predicted by the classical mixing-length theory of convection, especially in the outer atmospheric layers. These differences in the velocity profiles alone may lead to differences in the predicted line strengths, especially for the strongest lines which form in the outer atmosphere.

\begin{figure*}[t!]
\resizebox{\hsize}{!}{\includegraphics[clip=true]{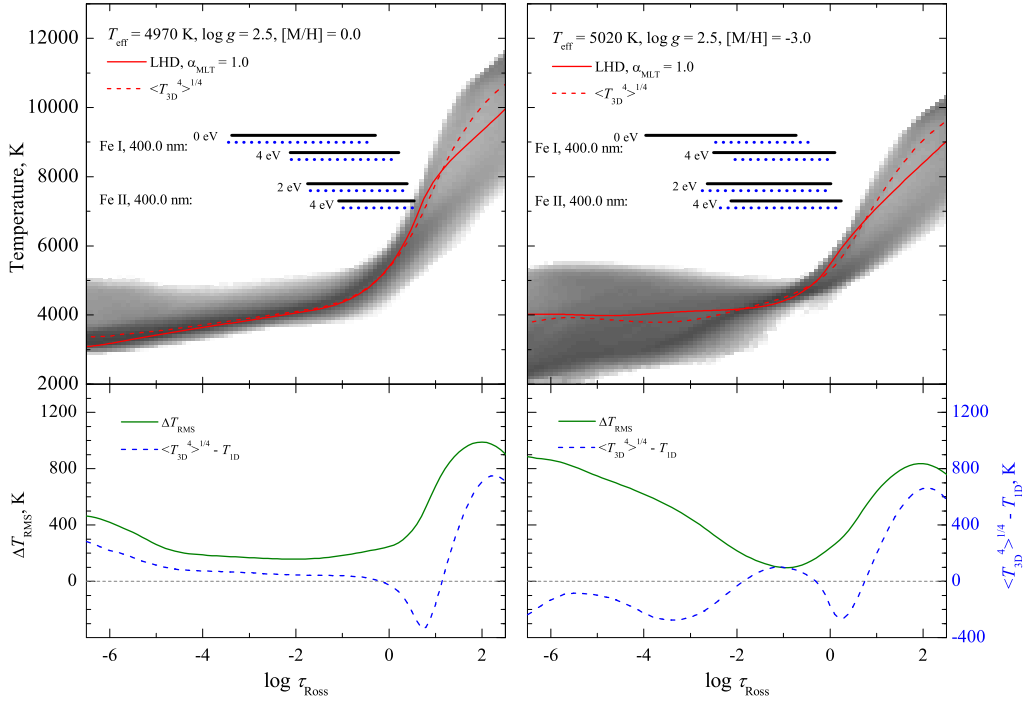}}
\caption{\footnotesize
\textbf{Upper panels}: temperature profiles in the 3D hydrodynamical (grey scale probability density plots), average $\xtmean{\mbox{3D}}$ (dashed red lines) and classical 1D (solid red lines) model atmospheres of a red giant ($\Teff\approx5000$\,K, $\log g=2.5$), shown at $\moh=0.0$ (left) and $\moh=-3.0$ (right). Horizontal bars indicate the approximate formation regions of weak \ion{Fe}{I} and \ion{Fe}{II} lines (equivalent width $W<0.5$\,pm), characterized by different excitation potentials (solid bars: 3D models, dotted bars: 1D models). \textbf{Lower panels}: profiles of horizontal RMS temperature fluctuations in the 3D models (solid green lines) and differences between the temperature profiles of the average $\xtmean{\mbox{3D}}$ and 1D model atmospheres (blue dashed lines; horizontal RMS temperature fluctuations were calculated as $\Delta T_{\rm RMS} = \sqrt{\langle(T - T_0)^2\rangle_{x,y,t}}$, where angled brackets denote temporal and horizontal averaging on surfaces of equal optical depth, and $T_0=\langle T \rangle_{x,y,t}$ is the average temperature at the given optical depth). In both panels, quantities related to the full 3D and average $\xtmean{\mbox{3D}}$ models were obtained using twenty 3D model snapshots which, in the case of $\xtmean{\mbox{3D}}$ models, were averaged on surfaces of equal Rosseland optical depths.
}
\label{fig:t-prof}
\end{figure*}

One common property of the 3D hydrodynamical model atmospheres of red giants is the existence of horizontal fluctuations of thermodynamical quantities seen at various geometrical and/or optical depths. In case of temperature, the amplitude of these fluctuations is different at different optical depths (Fig.~\ref{fig:t-prof}). Deep in the atmosphere, fluctuations tend to be large but they quickly decrease and approach a minimum close to the optical depth unity. From there on, fluctuations tend to monotonically increase again towards the outer atmospheric layers. Such behavior is related with the change of vertical and horizontal velocities with depth: large fluctuations at a given optical (or geometrical) depth below the optical surface are caused by different temperatures in the hotter up-flows and cooler down-flows in the granules and intergranular lanes, respectively. When the cooling outwardly-directed granular flows reach the optical surface, they are gradually deflected sideways which leads to significantly more homogeneous temperatures at this optical depth and thus, significantly smaller horizontal temperature fluctuations. However, the uprising material partly overshoots into the higher atmospheric layers and wave activity then takes over in the outer atmosphere, which produces larger horizontal temperature fluctuations again \citep[see, e.g.,][]{LK12}. Interestingly, such trends are qualitatively very similar in the giant models characterized by quite different effective temperatures, surface gravities, and metallicities, although the quantitative details (such as the amplitude of horizontal temperature fluctuations in the outer atmosphere) are of course different and depend on the atmospheric parameters, especially metallicity \citep[][]{CAT07,KSL13,DKS13}. The trends in horizontal temperature fluctuations seen in giants are also very similar to those seen in the models of dwarfs and subgiants, which points to the qualitatively similar patterns of convective motions in the atmospheres of these stars.

One should also note that typically there are significant differences between the temperature profiles of the average $\xtmean{\mbox{3D}}$ models and those of classical 1D model atmospheres, too (Fig.~\ref{fig:t-prof}). In case of 3D hydrodynamical \Stagger\ models of red giants, these differences may reach  $\sim1000$\,K in the outer atmosphere beyond $\log \tau_{5000}\lesssim-3.0$ \citep[][]{CAT07}. Our results obtained with the \COBOLD\ models do not show such large differences between the temperature profiles of the $\xtmean{\mbox{3D}}$ and 1D models but, qualitatively, predictions obtained with the \Stagger\ and \COBOLD\ model atmospheres are rather similar, in a sense that differences between the temperature profiles of the average $\xtmean{\mbox{3D}}$ and 1D model atmospheres tend to grow larger with decreasing metallicity.

\begin{figure*}[t!]
\resizebox{\hsize}{!}{\includegraphics[clip=true]{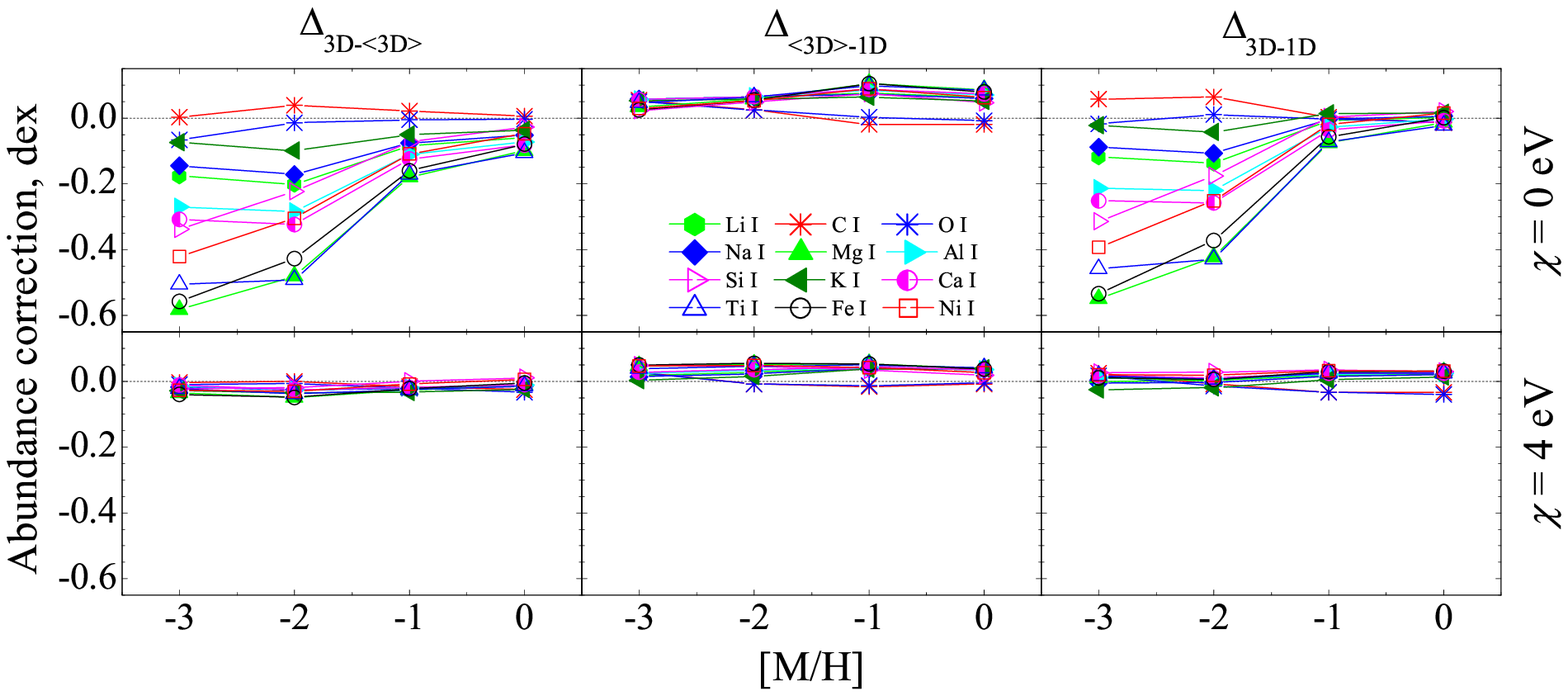}}
\caption{\footnotesize
Abundance corrections for weak spectral lines of neutral atoms at $\lambda=400$\,nm, computed using red giant model atmospheres at $\Teff\approx5000$\,K, $\log g=2.5$, and different metallicities. Three types of abundance corrections are shown: $\Delta_{\rm 3D-\langle3D\rangle}$ (left column), $\Delta_{\rm \langle3D\rangle-1D}$ (middle column), and $\Delta_{\rm 3D-1D}$ (right column). Corrections were computed for lines characterized with two excitation potentials, 0\,eV (upper row) and 4\,eV \citep[lower row; see][for further details]{DKS13}.
}
\label{fig:ac-atoms}
\end{figure*}

\subsection{The influence of convection on the spectral line formation in the atmospheres of red giant stars}

It is easy to anticipate from the discussion above that convection should have both direct and indirect influence on the formation of spectral lines in the atmospheres of red giant stars. Because of the large horizontal temperature and velocity fluctuations arising in the line forming regions due to convection (or convection-induced wave motions), local conditions for the spectral line formation may be strongly variable across the stellar atmosphere. Since temperature is one of the major factors determining line and continuum opacities, one may therefore expect that spectral line properties must be affected by the horizontal temperature fluctuations and differences between the temperature profiles of the average $\xtmean{\mbox{3D}}$ and classical 1D model atmospheres.

Indeed, differences between the line strengths predicted with the 3D hydrodynamical and 1D model atmospheres of red giants are significant. In case of weak artificial lines\footnote{We define artificial lines as those calculated with the arbitrary selected wavelength, excitation potential, and oscillator strength \citep[][]{SH02}. Such freedom allows to investigate the behavior of abundance corrections in a wide range of atomic parameters, and to study the influence of these parameters on the line formation properties \citep[see, e.g.,][]{CAT07,KSL13,DKS13}.} of neutral atoms, ionized atoms, and molecules (equivalent widths $W<0.5$\,pm), the strength of a given line depends on its atomic parameters (such as excitation potential and wavelength), ionization potential and dissociation energy of a given atom or molecule, respectively, as well as on the atmospheric parameters of the underlying model atmosphere. For the spectral lines of neutral atoms, $\Delta_{\rm 3D-1D}$ abundance corrections (i.e., differences between the abundances predicted by the 3D hydrodynamical and classical 1D model atmospheres) are very sensitive to the metallicity of a given model atmosphere, thus their absolute values may reach $\Delta_{\rm 3D-1D}\sim-0.6$\,dex at $\moh=-3.0$ (Fig.~\ref{fig:ac-atoms}). In the case of molecules, $\Delta_{\rm 3D-1D}$ corrections may become even larger, e.g., for CO it may attain $\Delta_{\rm 3D-1D}\sim-1.5$\,dex at $\moh=-3.0$.

Interestingly, the relative importance of the horizontal temperature fluctuations and differences between the temperature profiles of the average 3D and 1D model atmospheres depends on the metallicity, too. At higher metallicities, the abundance corrections due to horizontal temperature fluctuations, $\Delta_{\rm 3D-\langle3D\rangle}$, and corrections due to differences in the temperature profiles, $\Delta_{\rm \langle3D\rangle-1D}$\footnote{Since the average $\xtmean{\mbox{3D}}$ model does not retain information about the horizontal temperature fluctuations, the $\Delta_{\rm 3D-\langle3D\rangle}$ correction can be used to estimate the importance of horizontal temperature fluctuations. Similarly, the $\Delta_{\rm \langle3D\rangle-1D}$ correction arises due to differences between the temperature profiles of the average $\xtmean{\mbox{3D}}$ and 1D model atmospheres \citep[see, e.g.,][for details]{CLS11,KSL13,DKS13}.}, are nearly equal. However, horizontal temperature fluctuations start to dominate at lower metallicities, causing larger total  abundance corrections, $\Delta_{\rm 3D-1D}$ (Fig.~\ref{fig:ac-atoms}). Similar trends are observed in the case of ionized atoms, too \citep[][]{DKS13}. Obviously, the usage of the average $\xtmean{\mbox{3D}}$ models alone for estimating the differences between the predictions of the 3D and 1D models (as it has been repeatedly done in the past) may then be misleading. The real differences can only be assessed when full 3D models are used in the spectral abundance analysis.

It should be stressed that abundance corrections should in fact be different for stronger lines, which form over a larger range of optical depths and therefore should experience local temperatures and fluctuations that are different from those in the regions where the formation of weaker lines takes place. Additionally, stronger lines become sensitive to velocity fields, which in the case of 1D model atmospheres are accounted for by using depth-independent microturbulence velocity, $\xi_{\rm mic}$. In principle, abundance corrections inferred from the strictly differential 3D--1D analysis\footnote{Analysis done with the 3D and 1D models calculated using identical atmospheric parameters, equation of state, opacities, chemical composition, and numerical methods as far as possible.} should be insensitive to the error in the value of microturbulence used. For example, if the derived microturbulence velocity would be smaller than its ``real'' value, the 3D--1D abundance correction will be correspondingly larger than the one expected for the ``correct'' value of $\xi_{\rm mic}$ (and vice-versa), which would compensate for the ``incorrectly'' determined microturbulence velocity. However, such reasoning implies that current 3D hydrodynamical models are capable to realistically reproduce velocity fields and velocity fluctuations in stellar atmospheres. This, unfortunately, may not yet necessarily be the case: for example, current analysis of \ion{Fe}{i} line formation with the 3D hydrodynamical \COBOLD\ model of Procyon may indicate that a certain amount of small-scale velocity fluctuations may still be missing in the current 3D hydrodynamical model atmospheres \citep[][see also Steffen et al., this volume]{SLC09}. Obviously, further work is needed to make progress with the modeling of smaller-scale turbulent motions in the 3D hydrodynamical model atmospheres.

\begin{figure*}[t!]
\resizebox{\hsize}{!}{\includegraphics[width=6cm]{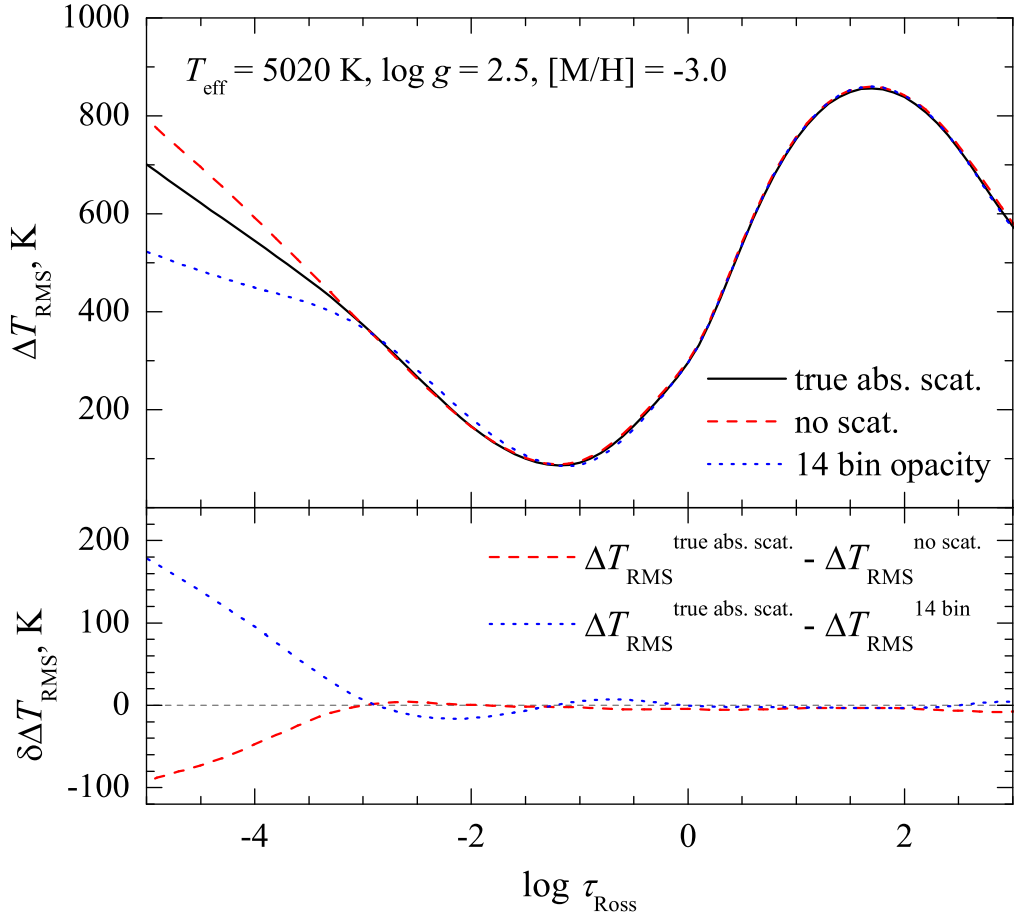}}
\caption{\footnotesize
\textbf{Upper panel}: profiles of horizontal RMS temperature fluctuations, $\Delta T_{\rm RMS}$, in the 3D hydrodynamical models of red giants ($\Teff=5020$\,K, $\log g=2.5$, $\moh=-3.0$), calculated with different assumptions regarding radiative transfer and plotted versus Rosseland optical depth. Solid (black) line shows RMS temperature fluctuations in the standard 3D hydrodynamical model, the dashed (red) line is $\Delta T_{\rm RMS}$ profile in the model where scattering is accounted for in an approximate way, and the dotted (blue) line is the RMS temperature profile in the model calculated using 14-bin opacities and scattering treated as true absorbtion. \textbf{Lower panel}: differences between the RMS temperature profiles shown in the upper panel.
}
\label{fig:t-scat}
\end{figure*}

\subsection{3D hydrodynamical modeling of red giant atmospheres: some problems and future developments}

Despite continuous improvement in the CPU architecture and increase in the CPU speed, calculation of the 3D hydrodynamical model atmospheres still remains very expensive in terms of CPU time. This is especially relevant in case of red giant stars, mostly because in the atmospheres of red giants the ratio of the radiative time scale to the Courant-Friedrich-Levy time decreases towards the upper RGB, leading to longer wall-clock times needed to compute the red giant models \citep[see also][]{LCS09}. Often the problem is made computationally more tractable by utilizing certain simplifications/adjustments related with the model physics, model setup, numerical issues, and so forth.

For example, to simplify the solution of radiative transfer problem, scattering is treated as true absorbtion in the standard \COBOLD\ setup. This approach has been recently questioned by \citet[][]{CHA11} and \citet[][]{HAC11} who found that thermal structures of the 3D hydrodynamical models calculated with scattering treated as true absorbtion may be very different from those computed utilizing a full treatment of coherent isotropic scattering. These findings led these authors to argue that differences in the thermal structures obtained with the \Stagger\ and \COBOLD\ codes may in fact be due to the different treatment of scattering. The latter claim has been recently questioned by \citet[][]{LS12} who found that differences between the average temperature profiles corresponding to the standard \COBOLD\ model and the model in which the contribution of scattering is treated in an approximate way\footnote{It has been demonstrated by \citet[][]{CHA11} and \citet[][]{HAC11} that thermal structures of the \Stagger\ models computed with coherent isotropic scattering were very similar to those where scattering opacity was ignored in the optically thin atmospheric layers.} are in fact significantly smaller than differences obtained by \citet[][$\sim100$\,K versus $\sim600$\,K at $\log \tauross\sim-4.0$, respectively]{CHA11}. \citet[][]{LS12} suggested that these differences could at least partly be due to differences in the calculation of binned opacities used in the \Stagger\ and \COBOLD\ codes.

To check the importance of opacity binning in the calculation of 3D hydrodynamical model atmospheres, we computed \COBOLD\ model of a red giant at $\Teff=5020$\,K, $\log g=2.5$, $\moh=-3.0$, using monochromatic \MARCS\ opacities grouped into 14 opacity bins (the standard \COBOLD\ model with the same atmospheric parameters was computed using 6-bin opacities). Indeed, we find that there are differences between the predictions of the 14-bin and 6-bin models (see Fig.~\ref{fig:t-scat}), but these differences ($\lesssim100$\,K) are significantly smaller than those obtained by \citet[][]{CHA11} for the \Stagger\ models computed with the different treatments of scattering. One may thus conclude that at least in the case of this particular \COBOLD\ model of the red giant the differences in the opacity binning scheme do not lead to significant differences in their thermal profiles. Nevertheless, both the opacity binning and scattering are important ingredients of the 3D hydrodynamical models, therefore their proper implementation into the 3D hydrodynamical stellar model atmosphere codes is clearly very important.

Strong convective motions in the atmospheres of red giant stars and short radiative timescales make the modeling of their atmospheres significantly more cumbersome than, e.g., those of dwarfs. For example, extremely large horizontal temperature gradients in the sub-photospheric layers often lead to situations where drops in the local temperature of several thousand Kelvin may occur over a few numerical grid points, which, in turn, may lead to significantly reduced computational time step and long model calculation time (or, in the worst case, to the model crash). Often the only viable solution is to increase the grid resolution but this takes its toll on the model size and the CPU time needed to compute the model. However, in the most extreme cases (e.g., low gravities and high effective temperatures) even such brute-force approach is sometimes not sufficient. Clearly, such issues may lead to serious complications when computing high-resolution model atmospheres and/or large grids of 3D hydrodynamical models. Therefore, further work on the improvement of numerical schemes for computing radiative transfer in the 3D hydrodynamical model atmosphere codes may be one of the priority tasks in the future development of the 3D model atmosphere and spectral synthesis codes.

\section{Conclusions}

We provide a brief overview of current progress in the modeling of red giant atmospheres with the 3D hydrodynamical model atmosphere code \COBOLD. The results obtained so far clearly indicate that convection plays significant role in the atmospheres of red giant stars, directly affecting their atmospheric structures and observable properties. Horizontal temperature fluctuations in the 3D hydrodynamical models, as well as differences between the temperature profiles of average $\xtmean{\mbox{3D}}$ and 1D model atmospheres, may lead to large discrepancies in the spectral line strengths predicted in 3D and 1D, especially at low metallicities ($\moh<-2.0$). Unfortunately, there may be no simple way to estimate the differences expected in, e.g., elemental abundances obtained with the 3D and 1D model atmospheres. Therefore, to properly account for the effects of convection, 3D hydrodynamical model atmospheres should be used whenever possible.

Despite the significant progress in the modeling of red giant atmospheres with the 3D hydrodynamical stellar atmosphere codes, simulations of red giant atmospheres still remain difficult and time consuming. To make the numerical problem more tractable, numerous simplifications with respect to the model physics, model setup, and numerical aspects of the model calculations are routinely applied in the calculation of the 3D hydrodynamical model atmospheres. However, physical properties of convective motions in the red giant atmospheres (such as extremely steep horizontal gradients of dynamical and thermodynamical quantities, short radiative time scales) will require further improvements of the numerical schemes utilized with the 3D hydrodynamical model atmosphere codes.

\begin{acknowledgements}

We thank the organizers of the 2nd \COBOLD\ workshop for a very successful and scientifically stimulating event, and for the financial assistance helping AK, JK, and DP to attend the event. AK thanks Sergei Andrievsky (Odessa observatory, Ukraine) for reading and commenting on the manuscript. This work was supported by grant from the Research Council of Lithuania (PRO-05/2012). HGL acknowledges financial support from EU contract MEXT-CT-2004-014265 (CIFIST), and by the Sonderforschungsbereich SFB\,881 ``The Milky Way System'' (subproject A4) of the German Research Foundation (DFG). AK and HGL acknowledge financial support from the the Sonderforschungsbereich SFB\,881 ``The Milky Way System'' (subproject A4) of the German Research Foundation (DFG) that allowed exchange visits between Vilnius and Heidelberg.

\end{acknowledgements}

\bibliographystyle{aa}

\end{document}